\documentclass[journal]{IEEEtran}
\usepackage{orcidlink}
\usepackage{amsmath,amsfonts}
\usepackage{algorithmic}
\usepackage{algorithm}
\usepackage{array}
\usepackage[caption=false,font=normalsize,labelfont=sf,textfont=sf]{subfig}
\usepackage{textcomp}
\usepackage{stfloats}
\usepackage{url}
\usepackage{verbatim}
\usepackage{graphicx}
\usepackage{cite}
\usepackage{booktabs}
\usepackage{float}
\usepackage{epsfig}
\usepackage{multirow}

\begin{document}

\title{Dielectric Barrier Corona Discharge Anomaly by Ionic Wind under Unipolar Voltage Excitation}

\author{Gan Fu\,\orcidlink{0000-0002-1825-0097} \\ g.fu@tue.nl}



\maketitle

\begin{abstract}
An anomalous back discharge movement phenomenon is induced by a set of dielectric barrier corona discharges (DBCD) at unipolar half-sine voltage waveforms, where the back discharge has a time delay that relates to the applied voltage level. An ionic wind model is employed to analyze the physical behavior. Theoretical explanation and quantitative analysis are presented in this study based on abundant experimental results of 5 typical insulating materials and a FEP insulating cable. A numerical model is derived, which indicates that the back discharge can be activated under a relatively low potential voltage level in this study. The results highlight that the back discharge movement phenomenon behaves distinctly under half-sine voltage with negative polarity, yielding a significantly different partial discharge (PD) pattern with positive polarity. Besides, PD amplitude dependent on dielectric thickness is demonstrated by plotting in phase resolved partial discharge (PRPD) pattern. Furthermore, comparative experiments are conducted with respect to the variation of air gap length and dielectric geometry, manifesting different influences on PD amplitude.
\end{abstract}

\begin{IEEEkeywords}
Partial discharge, Back discharge anomaly, DBCD, Ionic wind, PRPD pattern.
\end{IEEEkeywords}

\section{Introduction}
\IEEEPARstart{C}{orona} discharge is one of the most typical among various discharge types. It’s formed as a strong electric field exists between two inhomogeneous electrodes, yielding intensive discharges in partial location. The charges from the ionized air partially accumulate on the insulation surface while a dielectric material falls in between the electrodes, i.e. DBCD \cite{1timatkov_influence_2005,2yehia_characteristics_2019}. Enhanced PD detection methods are required as space charges drift in the air gap between electrodes, manifesting some abnormal electro-hydrodynamic (EHD) discharge phenomena \cite{3martins_modeling_2011}.

Partial discharge detection methods have received significant research attention as the evolution tools of insulation conditions, and the enhancement of experimental methods shows considerable analytical precision with respect to some specific discharge conditions \cite{1timatkov_influence_2005}. Recently, combined voltages have been popularly adopted in PD analysis \cite{4romano,5lindell,6hammarstrom,7hammarstrom_combination_2021,8wu,9guastavino,10wang,11suwarno_partial_1996,12mizutani2006pulse,13morsalin_corona_2018,14wang,15gui_partial_2020,fuPartialDischargeEvolution2022}. For instance, a superimposed sinusoidal AC voltage with DC offset was applied to PD qualification in HVDC system, namely ‘Direct Current Periodic (DCP)’ \cite{4romano}. Analytical experiments on square and semi-square voltages revealed the relation between PD extinction voltage and voltage rise time \cite{5lindell,6hammarstrom,7hammarstrom_combination_2021}, illustrated the transition process of PD amplitude with aging \cite{8wu}, and the PD behavior of material with void \cite{9guastavino}. The utilization of combined voltages is aimed at obtaining clearer and extra PD information in different discharge conditions. For instance, a unipolar half-sine waveform with pause time interval \cite{10wang}, as the voltage excitation, clarifies the influence of charge accumulation on PD activity. A periodic triangular voltage \cite{11suwarno_partial_1996,12mizutani2006pulse} was used to investigate PD behavior based on the method named ‘Pulse-Sequence Analysis (PSA)’, which provides an analogous PRPD pattern under the time domain. Furthermore, in terms of PD pattern recognition, sawtooth \cite{13morsalin_corona_2018}, trapezoidal and square waveforms \cite{14wang} were adopted for enhanced distinction on surface and cavity discharge. An exponential decay voltage was exerted on revealing the significant influence of on PD behavior in \cite{15gui_partial_2020}. Combined voltage waveform, especially unipolar voltage, yields similar space charge accumulation regularity as in DC voltage but also provides the ‘phase’ information by its AC component, making it possible for PRPD analysis with respect to space charge influence. 

In DC and unipolar conditions, the space charge accumulation effect should be taken into consideration. In 1920, Thomas Townsend Brown and Paul Alfred Biefeld found that a propulsion force is generated while high voltage is applied to an asymmetric copastor, namely the ’BB’ effect \cite{16article}. Many researchers attributed this effect to the ionic wind effect produced by corona discharge, and utilized it for electrostatic levitation \cite{3martins_modeling_2011,17soloviev_analytical_2012}, where the ionic wind is a kind of EHD gas flow originating from corona discharge \cite{18kitahara_experimental_2007}. The ionized charges in the drift region suffering the coulomb force follow the field direction, transmitting their momentum to air molecules, consequently forming an airflow. With the scenario that ionic wind will change the charge distribution on the insulation surface for DBCD models \cite{19yin_simulation_2016,20mehmood_analysis_2019,21shimizu_basic_2015}, experimental validation is required.

\begin{figure*}[!t]
\centering
\subfloat[]{\includegraphics[width=0.48\textwidth, height=6cm]{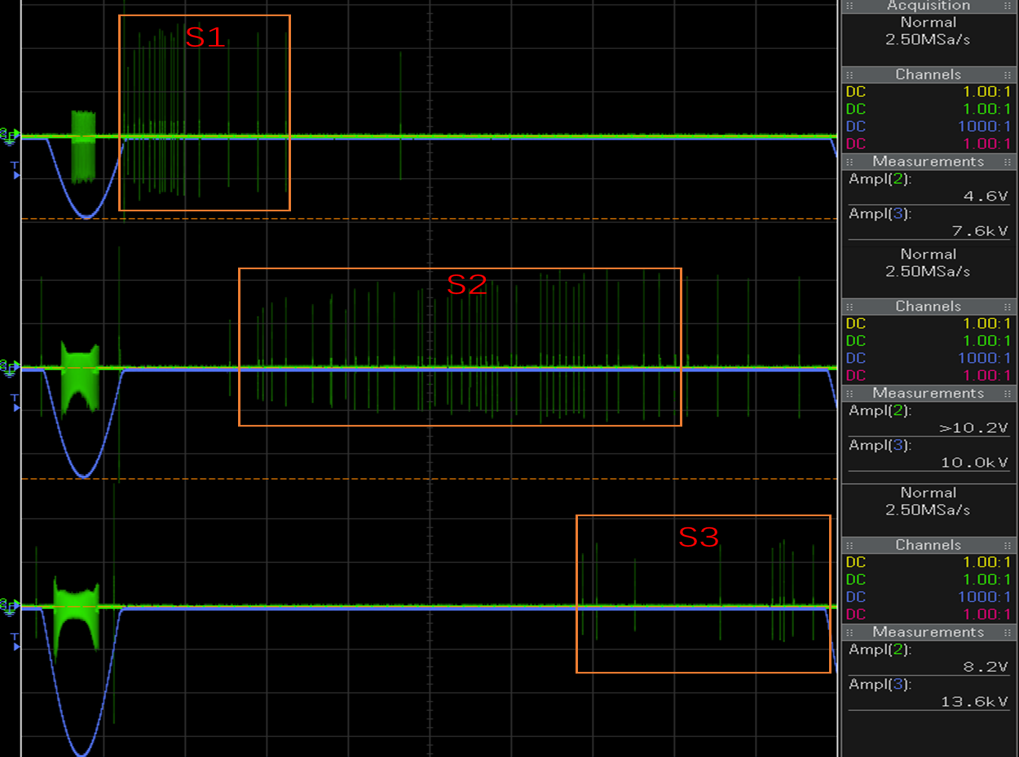}%
\label{fig_neg_case}}
\hfil
\subfloat[]{\includegraphics[width=0.48\textwidth, height=6cm]{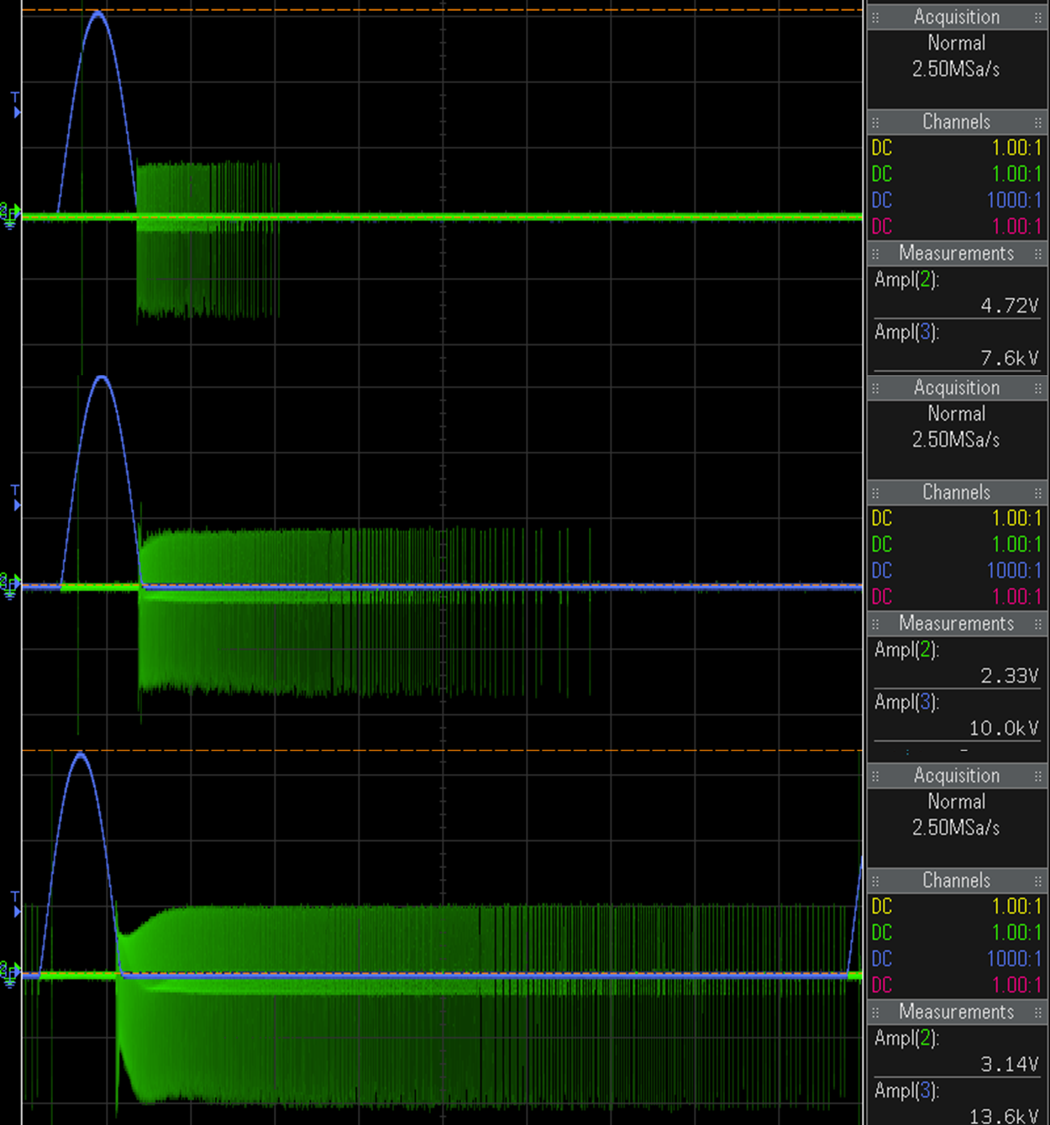}%
\label{fig_pos_case}}
\caption{Oscilloscope images of corona discharges under unipolar half-sine voltages. (a) Negative half-sine. (b) Positive half-sine.}
\label{fig_osc}
\end{figure*}

In this study, periodic unipolar half-sine voltages were utilized for investigating the PD activity influenced by charge accumulation on an insulating surface. A relaxation phase was introduced and the PD behavior during this time period was studied, which is introduced in Section. \ref{setction 2}. The initial measurement results indicate an anomalous movement of the back discharge concentration point under negative half-sine excitation, while under positive half-sine voltage is ‘normal’ as expected, as shown in Fig.\ref{fig_osc}. This behavior was further systematically investigated in Section. \ref{section 4}. Five different insulation materials and a FEP cable were tested to clarify this phenomenon under both positive and negative half-sine voltage waveforms. The dependence of PD amplitude on air gap distance, insulation thickness, and geometry are illustrated by PD patterns. Moreover, a numeric model is derived in Section. \ref{section 3} for the needle-plane geometry of this study.

\section{Experimental}
\label{setction 2}
In order to investigate the influence of space charge distribution on PD behavior during a zero-voltage "relaxation" phase, unipolar half-sine voltage waveforms are employed in this study. All the relevant experiments and discovered phenomena are based on the following time-resolved partial discharge measurement system.

\subsection{PD measurement system}
\begin{figure}[!h]
    \centering
    \includegraphics[width=0.48\textwidth]{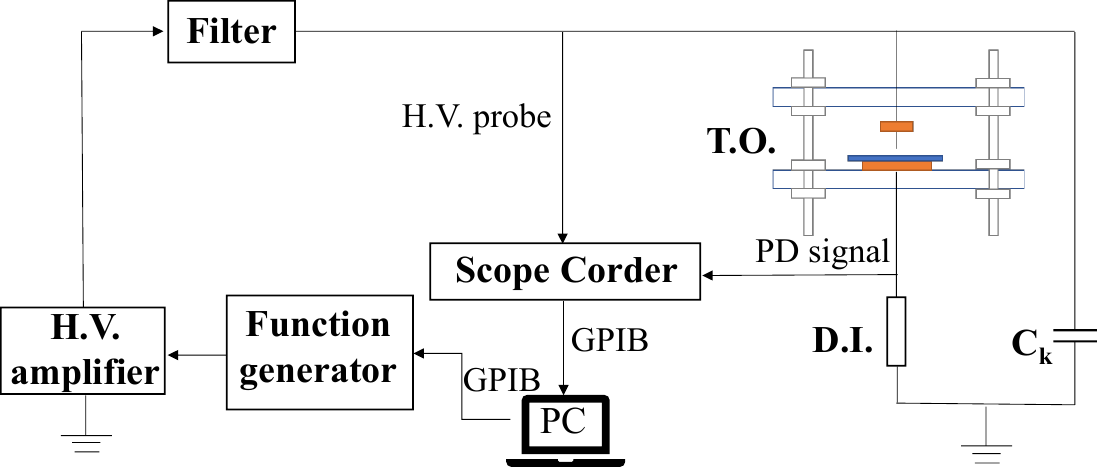}
    \caption{Diagram of time-resolved PD measurement system}
    \label{fig_system}
\end{figure}

The time-resolved partial discharge measurement system consists of the following elements: TREK  high voltage amplifier, HP 33120A function generator, Yokogawa DL750 Scope Corder, PMK PPE high voltage probe (1000:1), coupling capacitor and detection impedance, as shown in Fig. \ref{fig_system}.

The unipolar voltage waveforms were processed by using MATLAB, transmitted to and generated by the function generator, and subsequently amplified 2000 times through the TREK amplifier. The amplified voltages were submitted to the dielectric barrier needle-plane setup and generated partial discharges at a sufficiently high voltage magnitude. Detected PD current pulses were transformed into voltage signals and recorded by Scope Corder, which functioned as an oscilloscope with a resolution of 12 bits A/D. The high sampling rate and large deep memory of Scope Corder facilitate long-time recording.

\subsection{Half-sine voltage waveform}

The voltages employed in this study are unipolar half-sine waveforms as depicted in Figure 3. Each waveform consists of two distinct segments: a unipolar half of a sinusoidal waveform and a relaxation time period.
\begin{figure}[h]
    \centering
    \includegraphics[width=0.49\textwidth, height=0.22\textwidth]{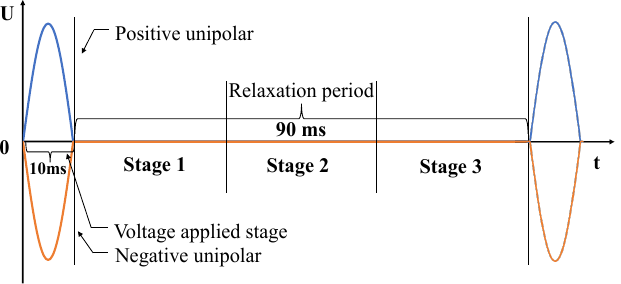}
    \caption{Unipolar half-sine waveform}
    \label{fig_waveform}
\end{figure}

In this study, the initial segment of the applied voltage waveform is a negative (or positive) half-period with a duration of 10 ms. The subsequent portion is characterized as the relaxation time period, with a voltage level of 0 V, providing, in these experiments, a 90 ms time interval for the movement and distribution of the surface charges. Notably, during this period, back discharges were observed to occur due to the accumulation of space charges on the insulation surface. For further analysis, the relaxation time period was divided into three distinct stages (S1, S2, S3) to indicate the time of occurrence of back discharges. Furthermore, the dependence of back discharges on voltage levels was investigated based on preliminary results.

\subsection{Sample cell and insulation materials}
Two kinds of corona discharge sample cells were induced for specific experimental investigations. The first one corresponds to a typical design of needle-plane configuration, commonly referred to as a DBCD setup. The other setup is specially tailored for needle-cable construction. Diagrams of these two sample cells are shown in Fig. \ref{fig_setup}:
\begin{figure}[!h]
\centering
\subfloat[]{\includegraphics[width=8.5cm, height=4.8cm]{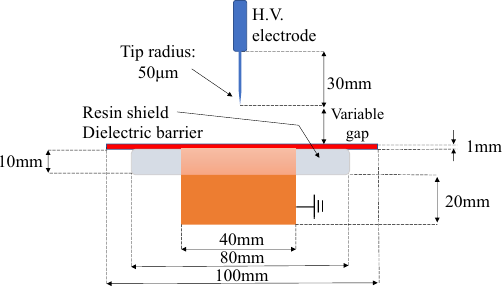}%
\label{fig_np_case}}

\subfloat[]{\includegraphics[width=8.5cm, height=3.7cm]{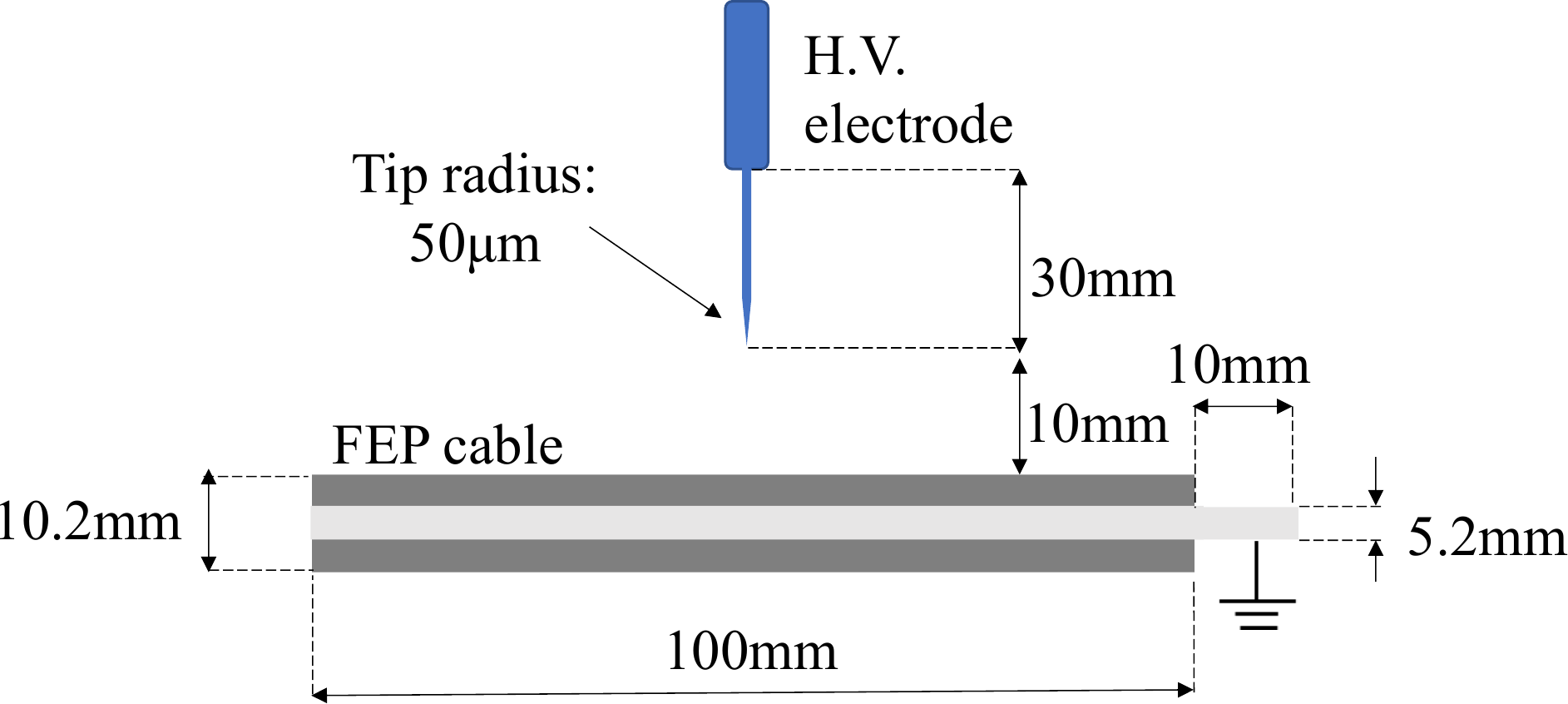}%
\label{fig_nc_case}}
\caption{Diagrams of sample cells of DBCD configuration.
(a) Needle-plane setup. (b) needle-cable setup.
}
\label{fig_setup}
\end{figure}
\begin{table*}[t!]
    \centering
    \caption{Bulk and surface resistivity measured at 400 V}
    \begin{tabular}{ccccccc}
    \toprule
         & PTFE & PE & PC & PVC & Pressboard & FEP cable\\
    \midrule
        Basic properties & nature, soft & nature, PE1000 & 
            \begin{tabular}{c}
                 transparent  \\
                 rigid
            \end{tabular} & nature, rigid & dry &  \begin{tabular}{c}
                 transparent  \\
                 rigid
            \end{tabular}\\ PDIV/kV & 4.2 & 4.0 & 3.7 & 3.9 & 2.7 & 4.4\\ Bulk resistivity/($\Omega\cdot$m) & $10^{16}$ & $4\times10^{15}$ & $7\times10^{14}$ & $5\times10^{14}$ & $7\times10^{10}$ & /\\ Surface resistivity/$\Omega$ & $8\times10^{14}$ & $5\times10^{14}$ & $3\times10^{12}$ & $2\times10^{12}$ & $4\times10^{13}$ & /\\
    \bottomrule
    \end{tabular}
    \label{tab_1}
\end{table*}

The ground electrode utilized in this study is a conductive brass cylinder embedded in epoxy resin, which has a diameter of 40 mm and thickness of 30 mm, and the resin shield, with a diameter of 80 mm and thickness of 10 mm. An insulation plate of diameter 100 mm and thickness 1 mm was positioned upon the ground electrode, maintaining the air gap distance as 10 mm. A needle electrode, measured 30 mm in length and 50 $\mu$m in tip radius, served as a high voltage electrode. The needle electrode was suspended by a brass holder with a banana socket, aligning to the center of the insulation plate. 

As for the needle-cable sample cell, analogous deployment was adopted, where the same needle electrode was suspended in the vertical centerline of the cable. The difference is the ground electrode (conductor layer) was embedded by cable insulation.

Six different insulation materials were utilised in this study: polytetrafluoroethylene (PTFE), polyethylene (PE), polycarbonate (PC), polyvinylchloride (PVC), unimpregnated pressboard and FEP cable. The former 5 materials were prepared into the same geometry configuration, while the cable set was selected to investigate the influence of dielectric geometry shape. Physical properties of the materials are shown in the Tab.\ref{tab_1}, where the bulk resistivity and surface resistivity were measured by KEITHLEY 6105 resistivity adapter and KEITHLEY 6514 electrometer at 400 V.

\subsection{Analysis method}
Each experimental trial comprised a sequence of eight consecutive cycles involving unipolar half-sine voltage pulses (10 ms), followed by a prolonged relaxation period of 90 ms. The applied voltage and PD pulse signals were recorded by Scope Corder with 10 MSa/s sampling rate simultaneously in the real-time domain. The acquired data encompasses both the PD pulse data and background noise. Furthermore, to facilitate further data analysis, a digital filtering process was employed to refine the PD signals.

The acquired data was stored in ‘*.WVF’ format, preserving the relative time of each PD event and its corresponding applied voltage. MATLAB programs were employed to read, cancel the noise, and extract PD pulses. PSA methods were adopted to visualize the time-dependent PD pulses through analogous PRPD patterns. Discharge voltage data was converted to PD amplitude (pC) using a PD calibrator, which yielded a specific coefficient of 15.5 pC/V.

\section{Mathematical model}
\label{section 3}
It is necessary to understand the relevant electrodynamic mechanism of the space charges in the needle-plane corona discharge anomalous phenomenon. Considering ionic wind theory in above-mentioned DBCD PD condition, as depicted in Fig.\ref{fig_ionic}. With negative excitation applied to the needle electrode, the strong electric field around the needle tip initiates the formation of an ionization region. The avalanche derived from the ionization region significantly increases the density of electrons, promotes the occurrence of electrons attachment to neutral air molecules, and consequently, generates the negative space charges. These negative charges are subjected to the Coulomb force, following the opposite direction as the electric field, which provide the force to propel charges from the ionization region into the drift region. During this movement, negative space charges collide with neutral air molecules and transfer their momentum to air molecules. The dynamic transfer of momentum ultimately forms air molecules flow towards the surface of insulation material \cite{19yin_simulation_2016}. The resulting airflow sweeps away the space charges on the insulation surface, particularly in the central area, to which the needle tip is aligned. As a consequence, a distinct charge valley forms in the center. In the meantime, with a large number of surface charges pushed outwards, a considerable proportion of surface charges are blown to the free space and dispersed in the air. Higher applied voltage induces stronger ionic wind, which forms a wider active ionic wind area, i.e. larger central charge valley. This type of valley or volcano shape distribution was also observed and measured in \cite{8785920}, which was illustrated by dielectric surface potential distribution.

In our case, the original PDs are generated by a negative half-sine voltage corona discharge, which causes negative charges deposition above the insulation surface, forming a strong back field. If take insulation surface as a voltage reference, the induced potential on the needle electrode will be positive, thus the back discharge generated around the needle tip is in positive polarity. Considering that the time constant of charge mobility on the dielectric surface with high surface resistivity is significantly larger than any single discharge process, as the charge mobility on the dielectric surface in this paper is tens of ms which is 6 orders larger than the single corona discharge duration, consequently, the back field for every single discharge that happens in the relaxation period can be regarded as electrostatic field condition.

Corona discharges manifest as non-equilibrium plasma characterized by a notably low level of ionization, (approximately 10$^{-8}$\%), which yield two distinct regions, namely the ionization zone and the drift zone \cite{22goldman_corona_1985}. The electric field strength can be given by:
\begin{figure*}
    \centering
    \includegraphics[width=0.98\textwidth]{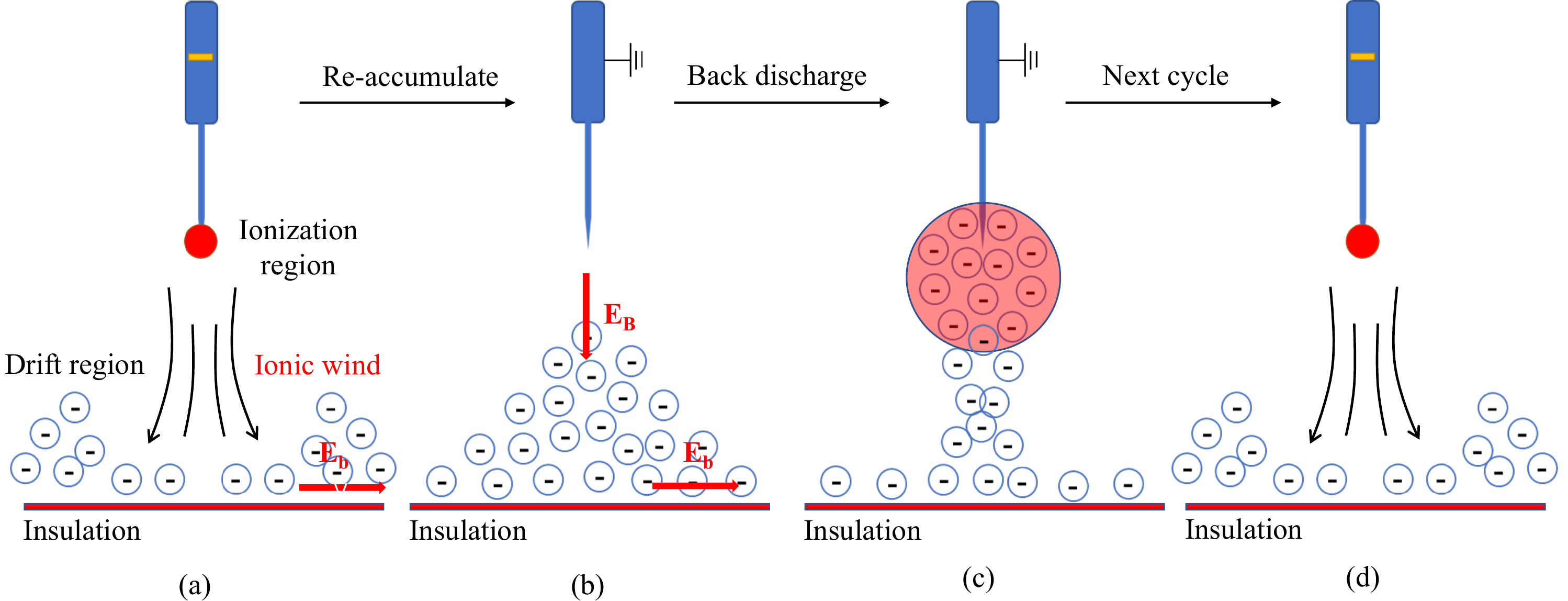}
    \caption{Phenomenology of ionic wind under negative half-sine voltage waveform}
    \label{fig_ionic}
\end{figure*}

\begin{equation}
    E=-\nabla V.
    \label{eq1}
\end{equation}
As known by Gauss’s law:$\nabla\cdot E=\rho_s /\varepsilon_0$, the potential difference $V$ around the needle can be obtained by solving the Poisson equation:
\begin{equation}
    \nabla^2 V=-\frac{\rho_s}{\varepsilon_0}.
    \label{eq2}
\end{equation}
Where $\rho_s$ is negative charge density in air gap space, as $\varepsilon_0$ is the permittivity in this region. 
The ionization region in this case is considered to be zero thickness, as suggested by Morrow \cite{24morrowTheoryPositiveGlow1997a}. And Kaptsov’s hypothesis \cite{3martins_modeling_2011} is introduced that, at the surface of the needle electrode, the electric field strength will increase proportional to the increased voltage, while it will maintain an approximately constant value after corona initiation. Then Peek’s empirical formula gives the inception voltage ($U_{0i}$) of the first corona discharge for the needle-plane geometry \cite{24fofana_study_2008,25beroual_discharge_2016}:
\begin{equation}
    U_{0i}=\frac{1}{2}R_NE_0\mathrm{log} \left (\frac{4D}{R_N}\right)\left(1+\frac{0.0436}{\sqrt{R_N}}\right).
    \label{eq12}
\end{equation}
Where $R_N$ is the needle electrode tip radius, $D$ is the air gap length, $E_0$ is the critical field indicating significant ionization phenomenon, as $E_0=31\delta(\mathrm{in\ kV/cm})$. Moreover, $\delta$ is the relative air density factor:
\begin{equation}
    \delta=\frac{P}{P_0}\frac{(273+T_0)}{(273+T)}=\frac{0.392P}{273+T}.
    \label{eq13}
\end{equation}
Where $P_0$ and $T_0$ are standard pressure ($760\ \mathrm{Torr}$) and temperature ($298.15\ \mathrm{K}$), $P$ and $T$ are ambient pressure and temperature in laboratory. $U_{0i}=0.365\ \mathrm{kV}$ is calculated for our case, namely, the minimum surface potential for generating back discharge upon the insulation surface. 

\section{Results and discussion}
\label{section 4}
\begin{figure*}[!t]
\centering
\includegraphics[width=0.33\textwidth, height=3.8cm]{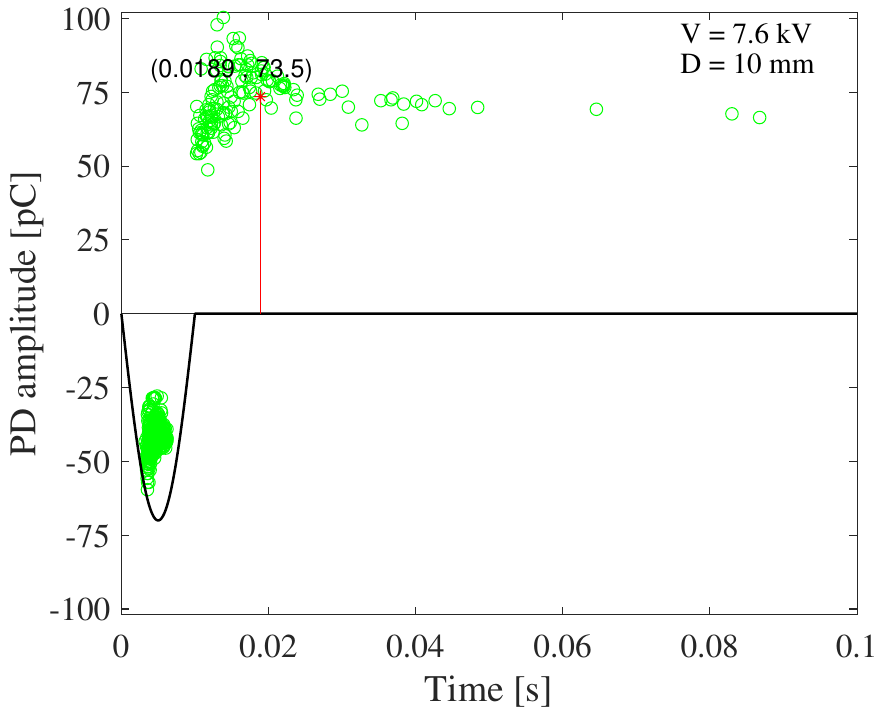}
\label{PC1}
\hfil
\subfloat[]{\includegraphics[width=0.33\textwidth, height=3.8cm]{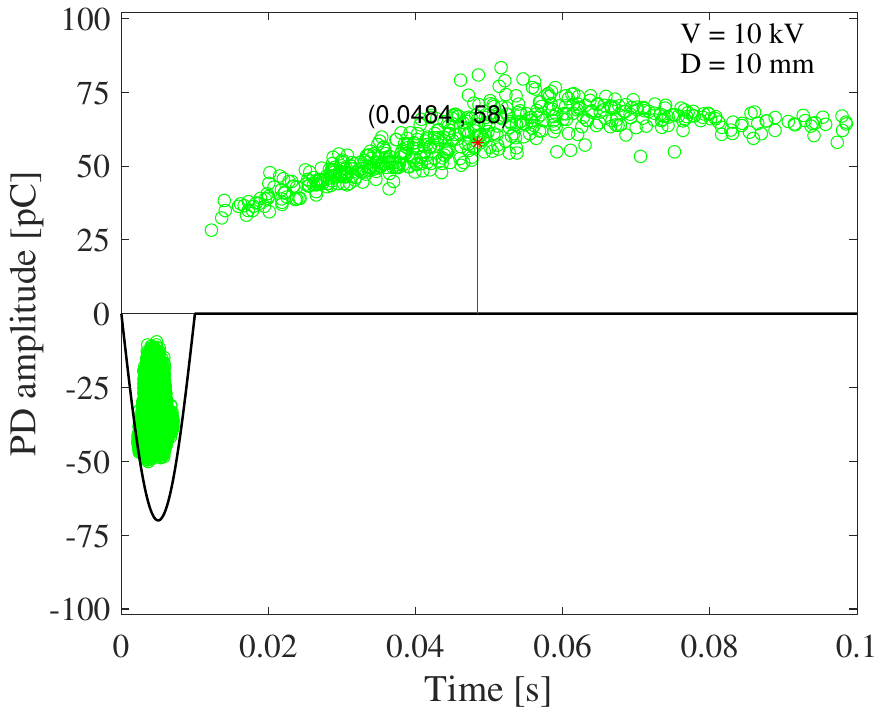}%
\label{PC2}}
\hfil
\includegraphics[width=0.33\textwidth, height=3.8cm]{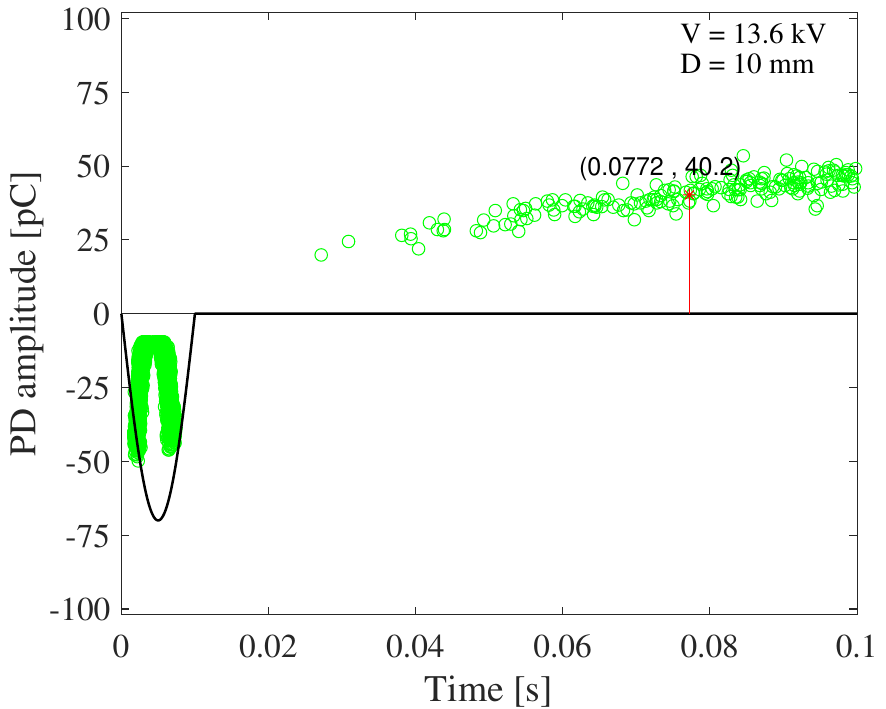}%
\label{PC3}\\
\includegraphics[width=0.33\textwidth, height=3.8cm]{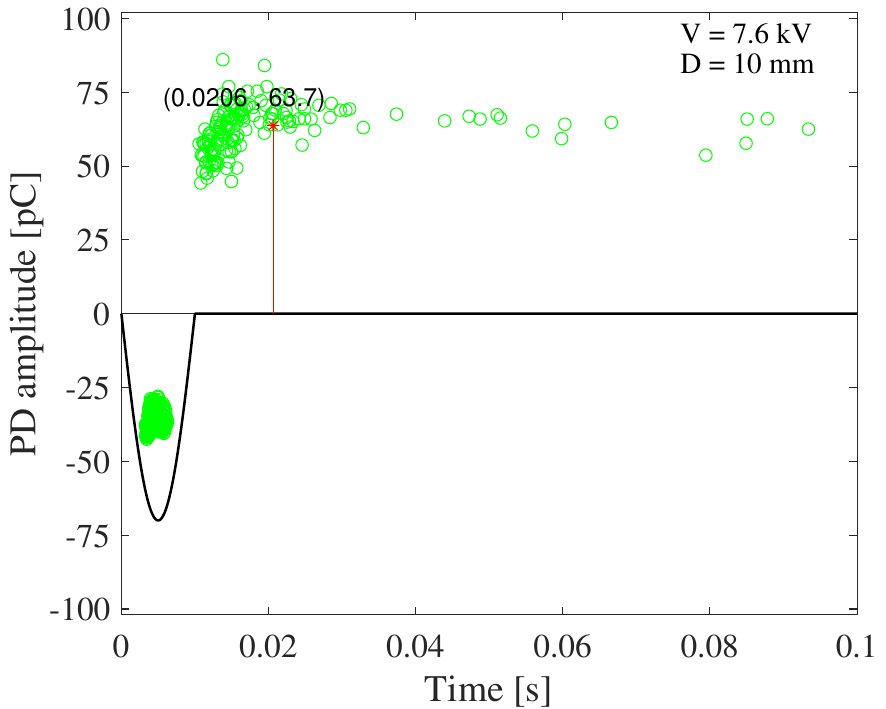}
\label{PVC1}
\hfil
\subfloat[]{\includegraphics[width=0.33\textwidth, height=3.8cm]{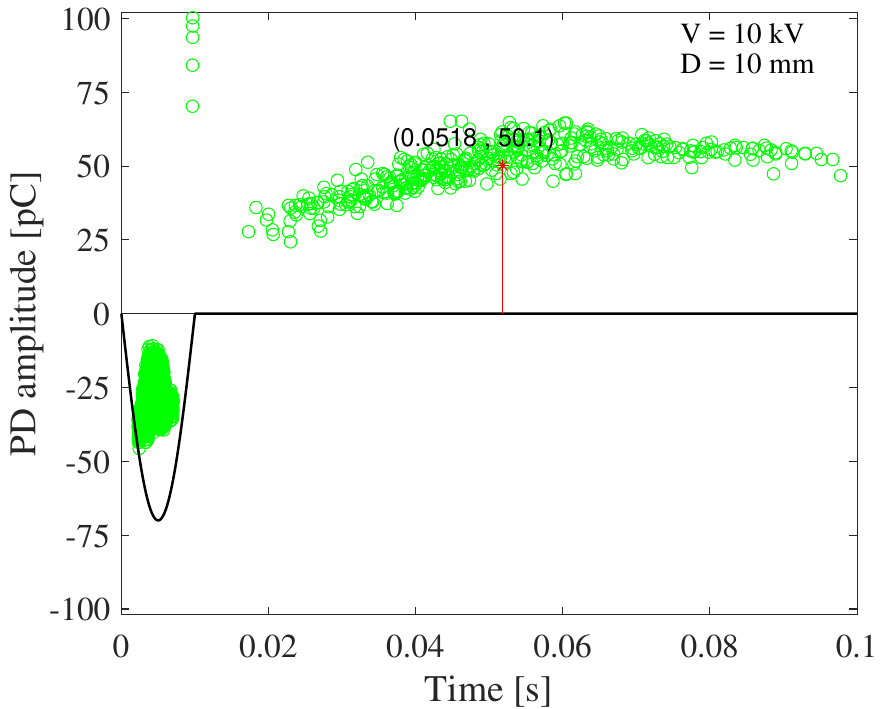}%
\label{PVC2}}
\hfil
\includegraphics[width=0.33\textwidth, height=3.8cm]{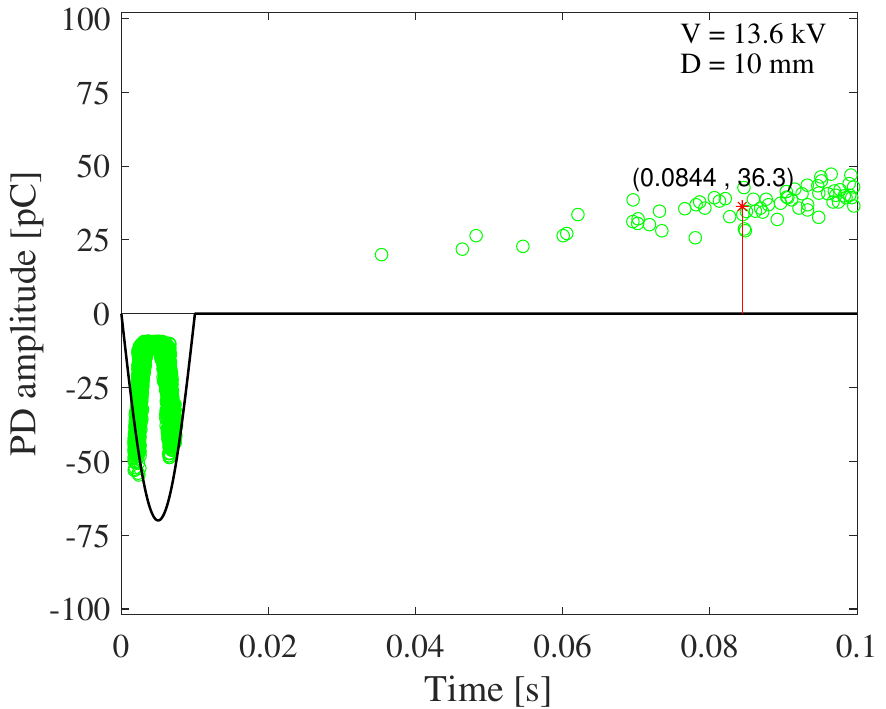}%
\label{PVC3}\\
\includegraphics[width=0.33\textwidth, height=3.8cm]{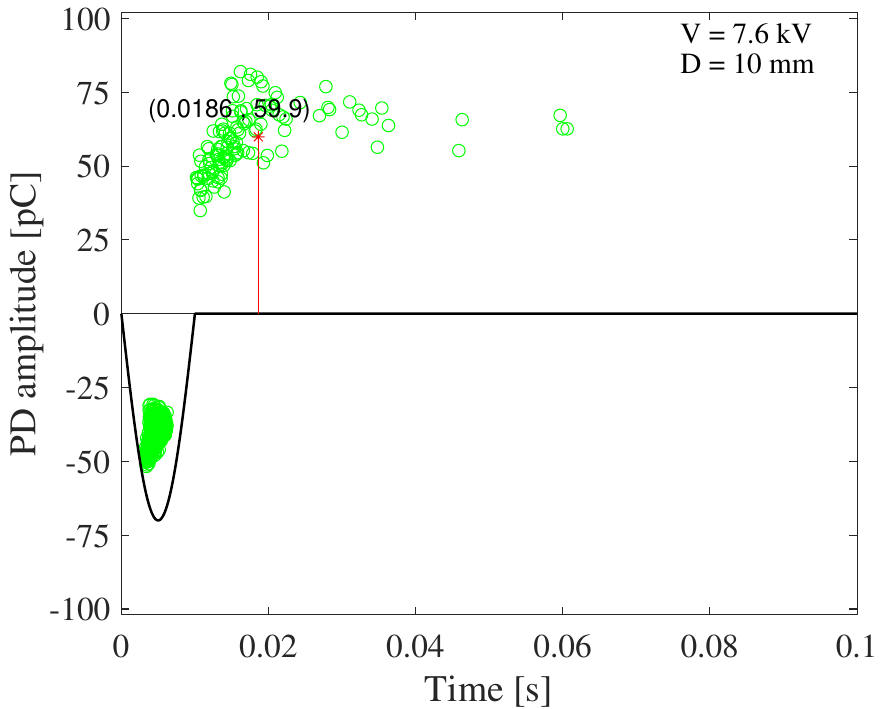}
\label{PE1}
\hfil
\subfloat[]{\includegraphics[width=0.33\textwidth, height=3.8cm]{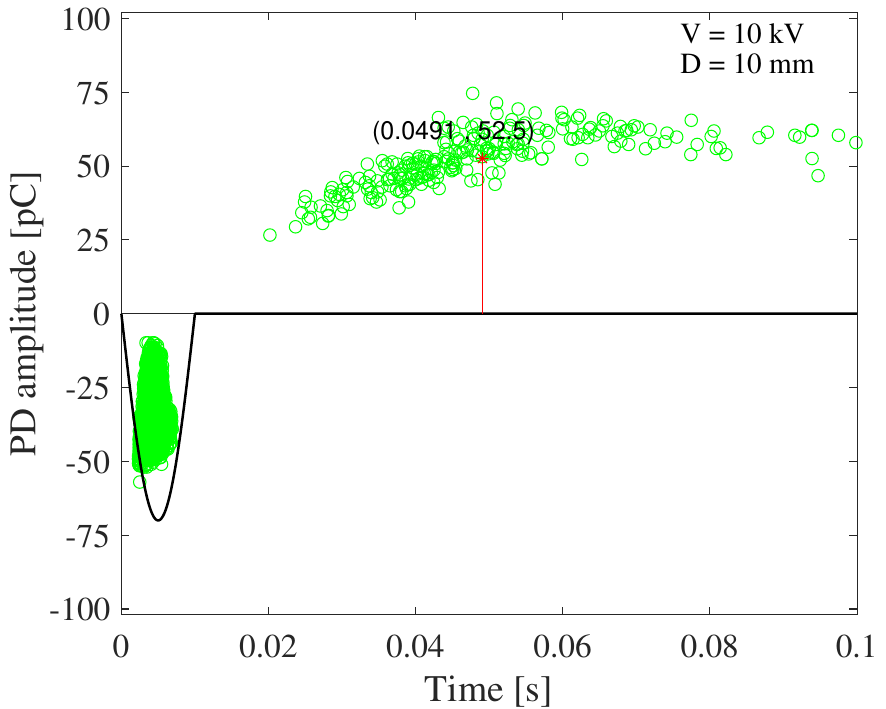}%
\label{PE2}}
\hfil
\includegraphics[width=0.33\textwidth, height=3.8cm]{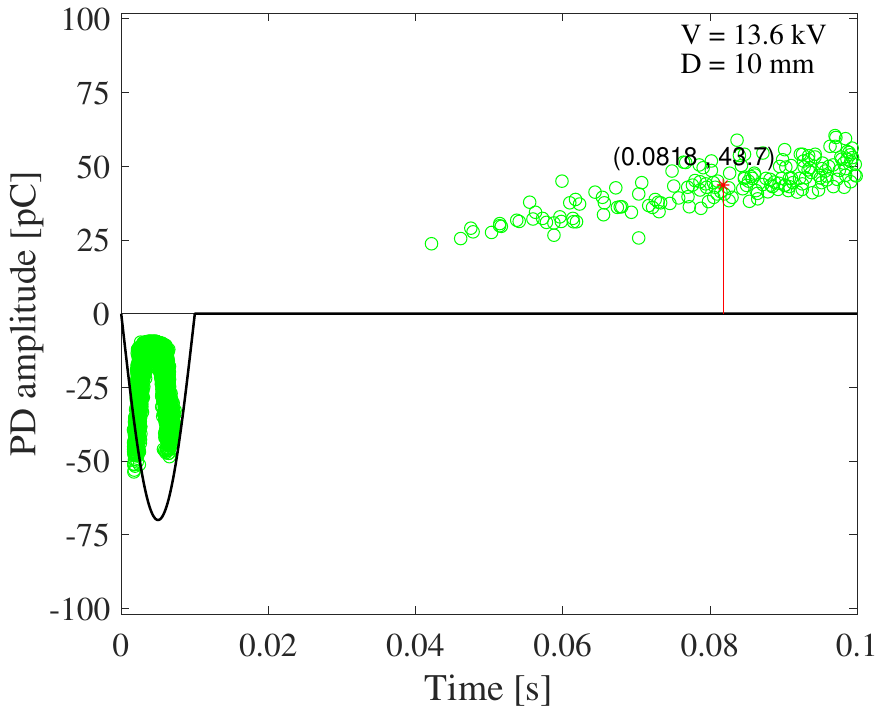}%
\label{PE3}\\
\includegraphics[width=0.33\textwidth, height=3.8cm]{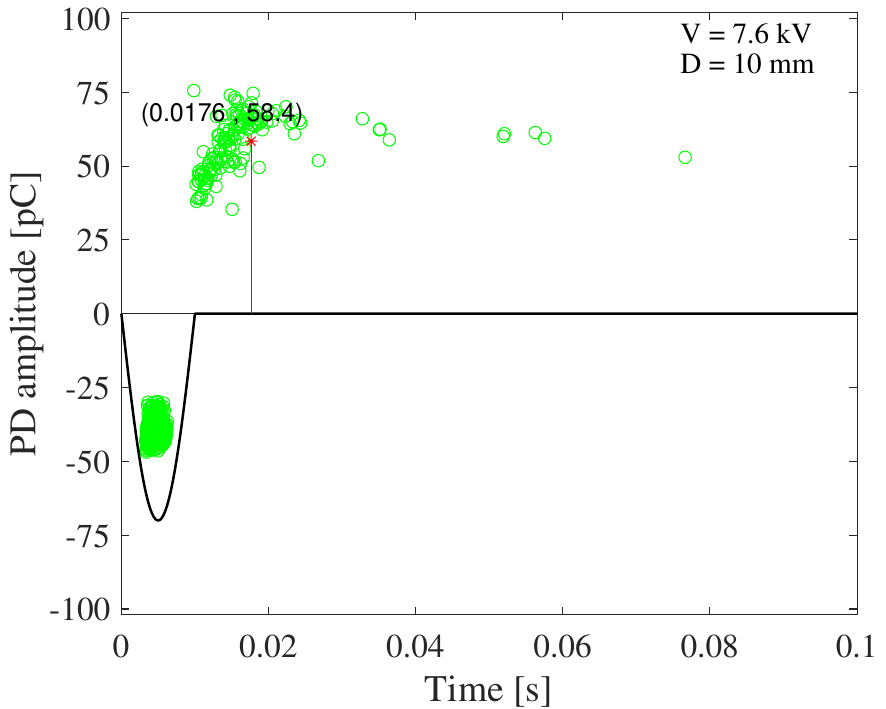}
\label{PTFE1}
\hfil
\subfloat[]{\includegraphics[width=0.33\textwidth, height=3.8cm]{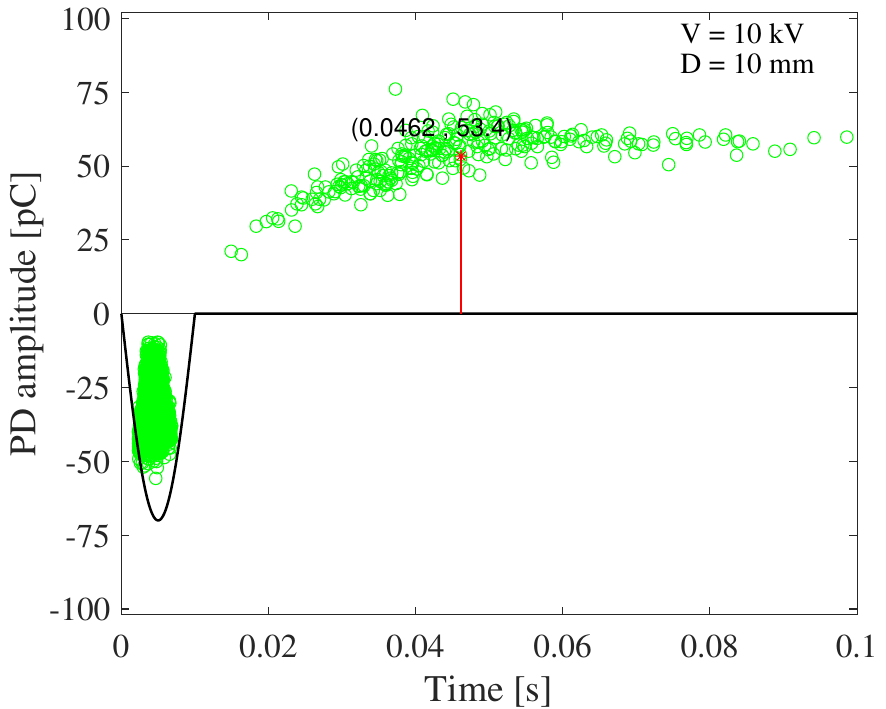}%
\label{PTFE2}}
\hfil
\includegraphics[width=0.33\textwidth, height=3.8cm]{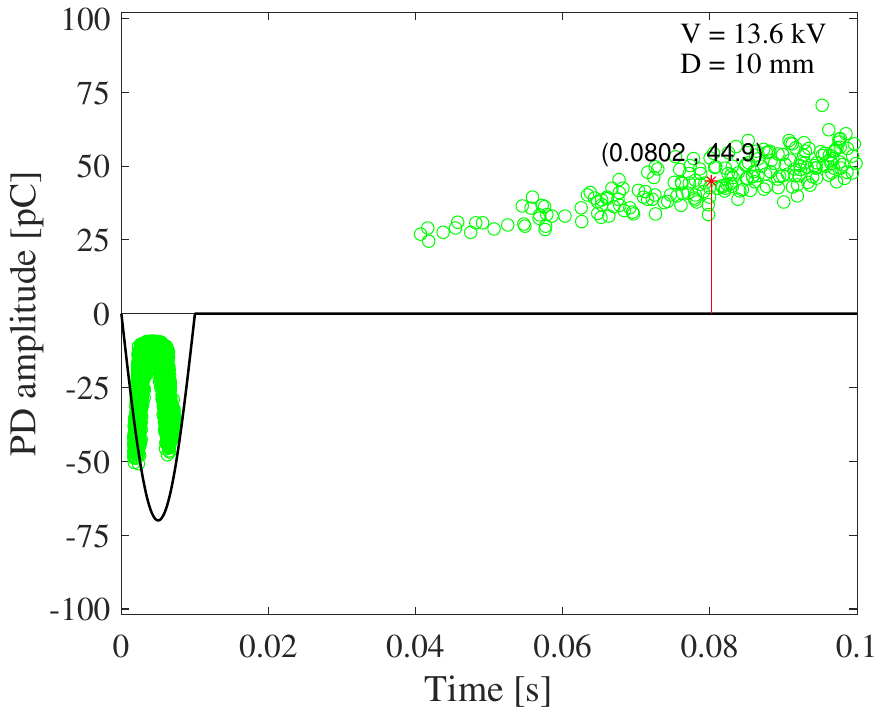}%
\label{PTFE3}\\
\includegraphics[width=0.33\textwidth, height=3.8cm]{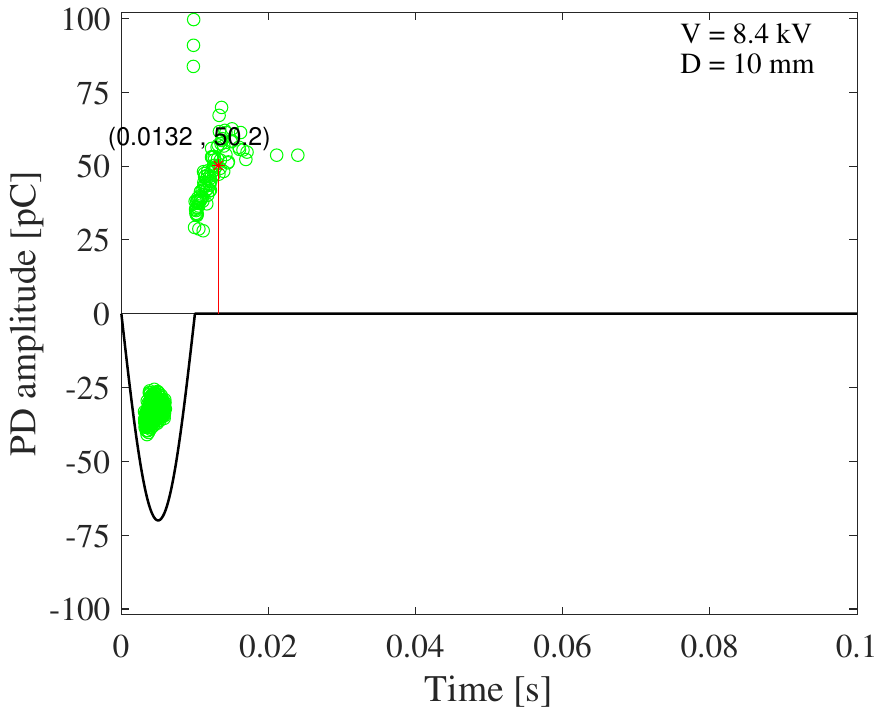}
\label{cable1}
\hfil
\subfloat[]{\includegraphics[width=0.33\textwidth, height=3.8cm]{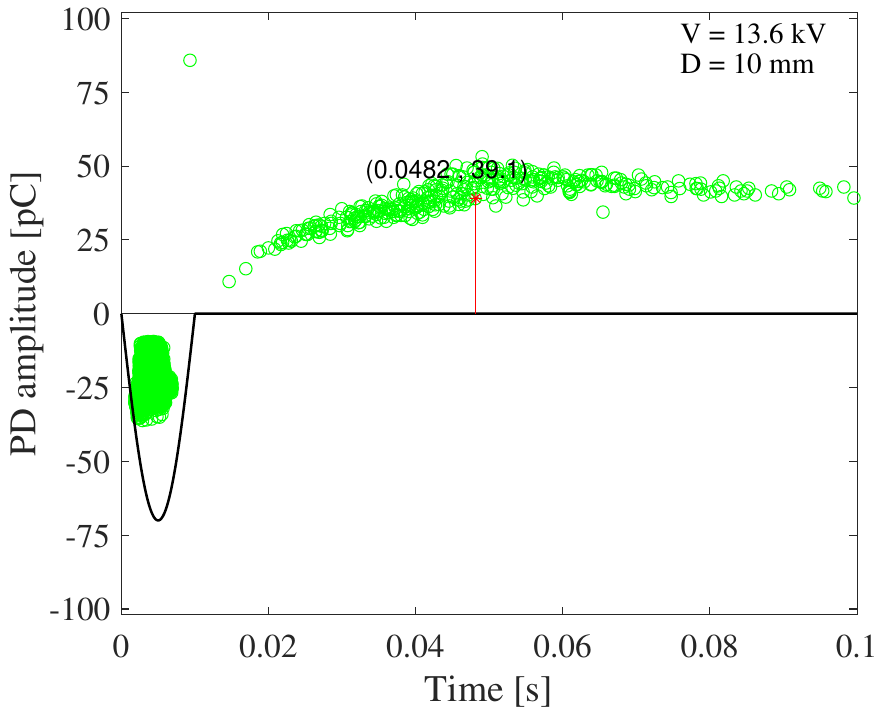}%
\label{cable2}}
\hfil
\includegraphics[width=0.33\textwidth, height=3.8cm]{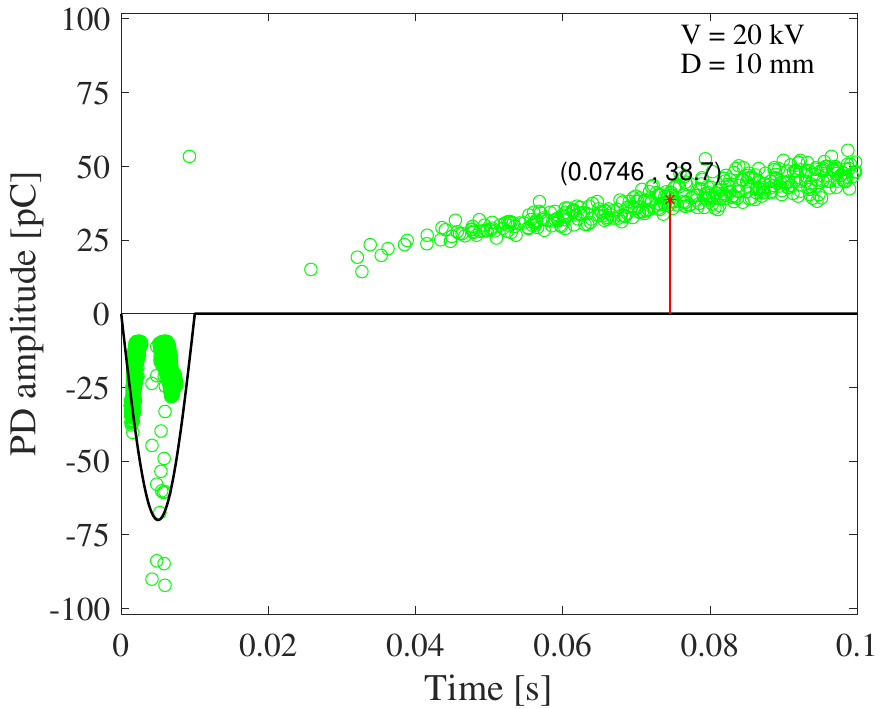}%
\label{cable3}\\
\caption{PD patterns of three stages for four materials: (a) PC. (b) PVC. (c) PE. (d) PTFE. (e) FEP cable.}
\label{fig_PD}
\end{figure*}

Measurement results from the six different insulation materials are presented in this section. The PD signals from the eight consecutive cycles were overlayed together into a single periodic for a more reliable PD pattern. A comprehensive analysis was conducted in this study, based on phase-resolved PD (PRPD) pattern. To accurately determine the occurrence of PD pulses within each cycle, the phase of applied voltage was substituted by the time of PD occurrence in each cycle.
A comparative overview of the examined materials is given in Tab.\ref{tab_1}. In addition to the pressboard, the partial discharge inception voltages (PDIV) of the other five materials are close. Consequently, the voltage levels that are required for the back discharge cluster to locate in different positions are also similar for these planar materials. In this study, 7.6 kV, 10 kV, and 13.6 kV were chosen and adopted as the trigger voltages of three back discharge stages, as shown in Fig.\ref{fig_neg_case} and Fig.\ref{fig_waveform}. PD patterns of each stage for each insulation material are shown in Fig.\ref{fig_PD}. Where V is the peak value of the applied voltage, and D is the air gap distance between the needle tip and insulation material surface. It is important to know that the red point with stem, namely the feature point, is an indicator of the average PD amplitude and occurrence time of PDs, which is summarized in Tab.\ref{Discharge_info}.

\begin{table}[t!]
    \centering
    \caption{Discharge information of five materials}
    \begin{tabular}[5cm]{ccccccc}
    \toprule
         \multirow{2}{4em}{Material}& 
         \multicolumn{3}{c}{\begin{tabular}{c}
             Avarage\\
             amplitude (pC)
         \end{tabular}}&
         \multicolumn{3}{c}{\begin{tabular}{c}
              Average\\
              position (ms) 
         \end{tabular}}\\
        
        &S1&S2&S3&S1&S2&S3\\
          \midrule
         PC&73.5&58&40.2&18.9&48.4&77.2\\
         PVC&63.7&50.1&36.3&20.6&51.8&84.4\\
         PE&59.9&52.5&43.7&18.6&49.1&81.8\\
         PTFE&58.4&53.4&44.9&17.6&46.2&80.2\\
         FEP cable&50.2&39.1&38.7&13.2&48.2&74.6\\
            \bottomrule
         \end{tabular} 
    \label{Discharge_info}
\end{table}
The investigation carried out in this study, as presented in Fig.\ref{fig_PD}, highlights the ’movement’ of back discharge clusters across all the above-mentioned materials, i.e. PC, PVC, PE, PTFE, and FEP cable. Some interesting regularities emerged from this analysis, which can be drawn as follows:

(1) With the increment of applied voltage, the region of back discharge concentration (indicated by feature point) moves forward along the relaxation time period and finally ‘disappears’ in the pattern with even further increased voltage. This phenomenon indicates that there will be no back discharge under repetitive negative half-sine voltage with sufficiently high amplitude and proper frequency. Furthermore, more experimental results have proven that this back discharge movement anomaly exists in various ranges of frequency, duty rate, and unipolar voltage construction, rather than only in a specific discharge configuration. 

(2) It can be observed that as the applied voltage increases, there is a gradual decreasing trend for the average amplitude of the back discharge cluster, which indicates an inverse relationship between voltage level and back discharge amplitude.

(3) The amplitude of back discharge varies across different materials, with a tendency for materials with higher surface resistivity to exhibit lower PD amplitudes, which suggests that the resistivity of materials also affects back discharge amplitude.

(4) Furthermore, the back discharge cluster in each S2 has a rising front with respect to PD amplitude. Which is an indicator of dielectric surface potential and charge density of the drift region.

Considering the participation of ionic wind theory, the ions generated by ionization are propelled towards the insulation surface and swept from the center of the insulation surface to the outer area during the ‘half-sinusoidal’ period, as depicted in Fig.\ref{fig_ionic}a. When shifting to the relaxation time period, the voltage potential of the needle electrode becomes 0 V. In this circumstance, outer charges on the insulation surface generate a horizontal back field $E_b$ along the insulation surface, promoting the propagation of inner charges towards the central surface, namely re-accumulation, as depicted in Fig.\ref{fig_ionic}b. The re-accumulated space charges in the central area ultimately lead to the occurrence of back discharge around the needle tip, as depicted in Fig.\ref{fig_ionic}c. Notably, the generation of a stronger ionic wind, driven by a higher voltage level, exerts a stronger ionic wind which results in further dispersion for charges. Consequently, charges take a longer duration to re-accumulate towards the center of insulation materials with high surface resistivity. This dynamic manifests in the PD pattern as the movement of the back discharge cluster. Furthermore, if further increase the applied voltage, the back discharge cluster continues moving along the phase of the relaxation time period until it eventually ‘disappears’ from the pattern. This behavior arises due to the fact that surface charges are blown too distantly to initiate a back discharge before the coming of the subsequent cycle, where the negative potential on the needle electrode will prevent the occurrence of back discharge. 

Additionally, higher applied voltage generates larger surface charge density, yielding a stronger back field, which endues higher drift velocity for a single charge in the drift region. Fewer charges are required for a single discharge process with higher charge drift velocity, leading to the decrement of discharge amplitude. As we can observe there is a rising front in each S2 in Fig.\ref{fig_PD}, which is because of, the exponential decay of surface potential \cite{26wang2013lumped}, leading to a gradual decrement of charge drift velocity. which requires more space charges for a single discharge.

It is worth noting that materials with higher surface resistivity exhibit lower back discharge amplitude at the same applied voltage. This observation can be attributed to easier attachment and more difficult dissipation characteristics of space charges for materials with higher surface resistivity, which maintain higher surface potential during the relaxation time period. Consequently, facilitates the formation of each single back discharge with lower space charge density around the needle tip, ultimately resulting in a lower back discharge amplitude.

\begin{figure}[h!]
    \centering
    \includegraphics[width=0.4\textwidth, height=0.3\textwidth]{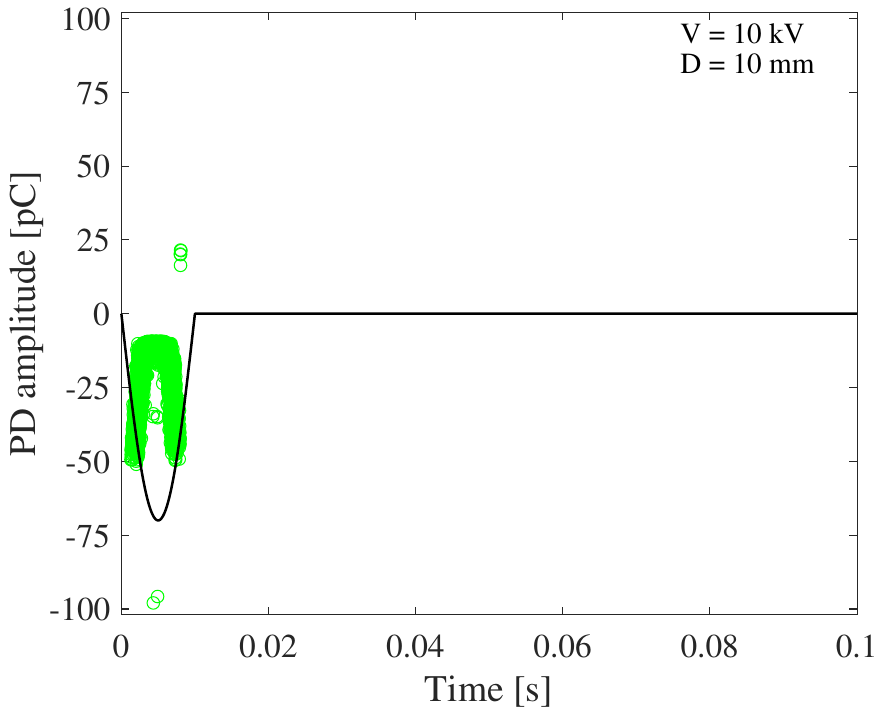}
    \caption{PD pattern of pressboard}
    \label{pressboard}
\end{figure}

Fig.\ref{pressboard} provides the PD pattern of the pressboard under 10 kV voltage level, exhibiting a different PD behavior compared to the other polymer-based insulation materials, i.e. entirely no back discharge happens during the relaxation time period.

The conductivity of unimpregnated pressboard, which inherently contains some moisture, is considerably higher compared to other materials. As a result, space charges tend to dissipate more readily cross through the pressboard, and inject into the ground electrode. Generally, there are several charge dissipation mechanisms, for instance, injection, neutralization, detrap, and re-distribution on dielectric surface, etc \cite{guanMechanismsSurfaceCharge2019}.
The back discharge movement behavior in this paper is explained through a first-order charge dissipation mechanism. Essentially, the predominant mechanism, characterizing the back discharge process, is assumed to be dominant over other secondary electrical-activated mechanisms. The onset of charge dynamics leads to different discharge phenomena around and within the insulation. Material with low bulk resistivity tends to yield no back discharge even with high surface resistivity in this study, for instance, the pressboard. This can be attributed to the charge injection mechanism coming to dominate, leading to the insufficient surface potential for back discharge inception. This PD behavior indicates that the back discharge cluster movement is primarily caused by space charge accumulation on the insulation surface.

For a better understanding of this back discharge behavior, extended partial discharge experiments were carried out, considering some other possible factors, namely, the thickness of insulation material, air gap length, voltage polarity, and insulation geometry.

\subsection{Influence of insulation material thickness}
\begin{figure}[h!]
    \centering
\includegraphics[width=0.48\textwidth, height=0.41\textwidth]{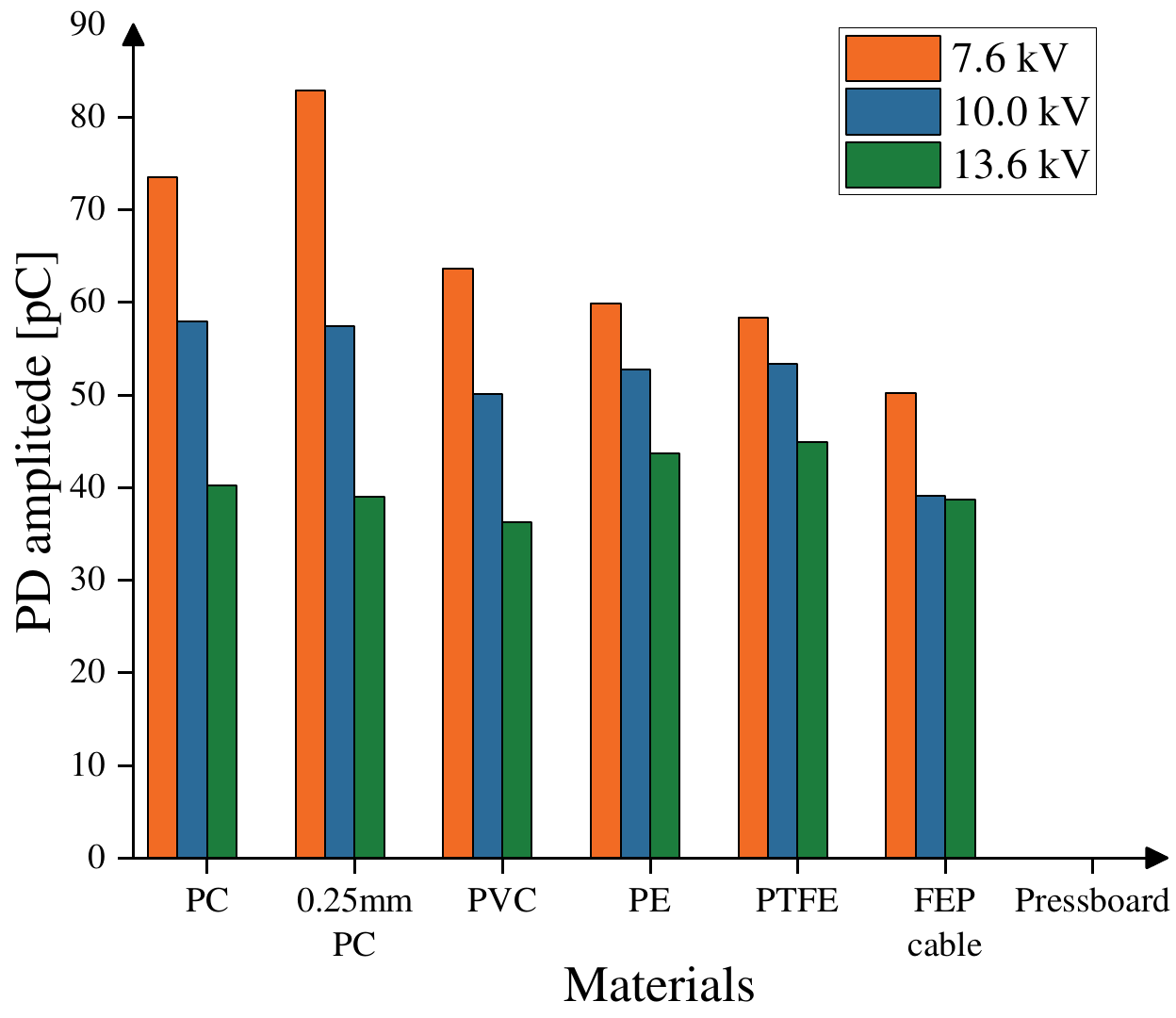}
    \caption{PD amplitude of different materials under three voltage levels}
    \label{PD_material}
\end{figure}

By utilizing a PC plate with a thickness of 0.25 mm under the same experimental deployment, the corresponding PD pattern shows the same back discharge trend, while the voltage levels for achieving each stage (S1, S2, S3) are significantly lower than thicker material, which are 7.2 kV, 9.2 kV, 12 kV respectively. Besides, the back discharge amplitude is higher than that of 1 mm thickness material in S1, as shown in the histogram in Fig.\ref{PD_material}.

As it's known that thinner material has higher capacitance, which leads to lower surface potential compared to thicker material if we consider the insulation plate as an ideal plate capacitor, which gives: $C=Q/U$. Whereas the net electric field can be represented by the difference between the main field and back field: $E_n=E_m-E_b$. The back field of thinner material is weaker, which explains why the required applied voltage level for thinner material to generate sufficient net electric field is lower.
Also, during back discharge occurrence, a larger amount of free charges are required for a single discharge in thinner material, leading to a higher average back discharge amplitude. 
\subsection{Influence of air gap length}
\begin{figure}[h!]
    \centering
\includegraphics[width=0.46\textwidth, height=0.35\textwidth]{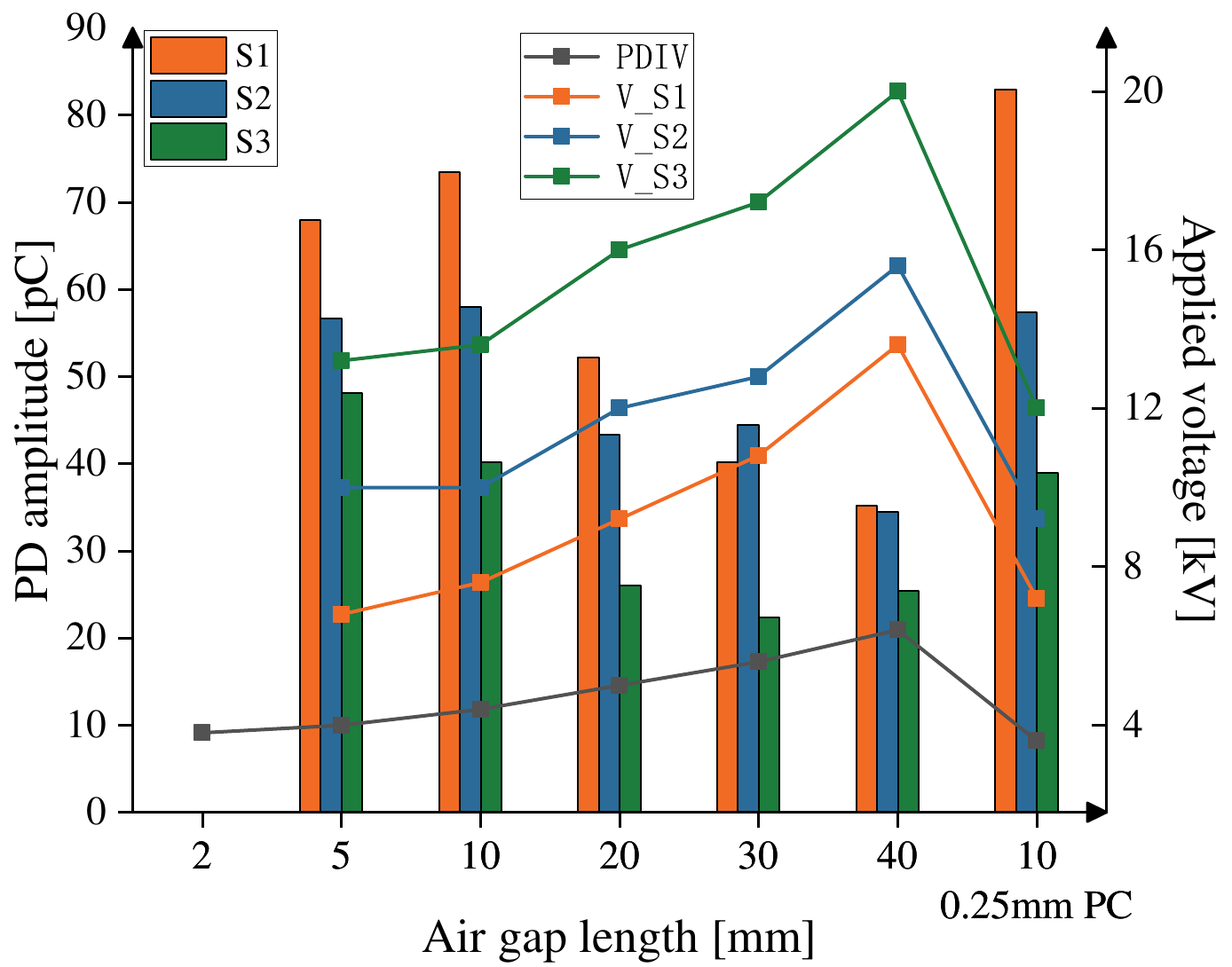}
    \caption{PD amplitude of different materials under three voltage levels}
    \label{PD_length}
\end{figure}

To investigate the influence of the air gap length on the PD amplitude of the back discharge cluster, a series of experiments were conducted by using a 1mm PC material under different air gap lengths, from 2 mm to 40 mm. The trigger voltages for each stage corresponding to different air gap lengths were depicted in Fig.\ref{PD_length}. Observation from the result indicates that, generally, larger needle-plane distance requires a higher trigger voltage level for each stage.

Furthermore, it is observed that the PD amplitude gradually decreases as the air gap length increases. This can be attributed to the combined effect of ionic wind and surface potential. It's reasonable to consider that lower charge density among the drift region and higher surface potential is derived from higher applied voltage for larger air gap distance, yielding a decreasing tendency of back discharge amplitude. However, experimental verification is not available due to the limitation of devices and techniques for surface potential and charge distribution measuring in this study.

Additionally, when the air gap is small enough, for instance, 2 mm, as shown in Fig.\ref{PD_length}, there is no back discharge cluster during the relaxation time period, substituted by a bunch of intensive back discharge around zero cross point. This is due to another discharge mechanism comes to the predominance, namely, streamer discharge. In such a case, the ionic wind has no space for propagation and transferring the momentum, whereas it’s much easier to generate a conductive path, i.e. streamer. Consequently, the charges in the space entirely drift into the streamer path at that moment. 

\subsection{Applied voltage polarity}
It is important to note that this back discharge movement doesn’t occur under positive half-sine voltage excitation as shown in Fig.\ref{fig_pos_case}, where the back discharges during the relaxation time period are negative polarity. It can be apparently observed that as the voltage increases, the occurrence of back discharge becomes more pronounced, following an exponential decay trend with respect to PD repetition rate throughout the relaxation time period. Notably, the discharges happen in the relaxation time period by positive excitation, predominantly generated by the ionization of air molecules rather than the drifting of free charges above the insulation surface which takes a significantly longer time to re-accumulate. Somehow, it's a clear illustration for understanding different discharge mechanisms of unipolar voltages with different polarities.

\subsection{Influence of geometry}
Analogous experiments were performed with the needle-cable setup, in order to investigate the influence of dielectric geometry, as shown in Fig.\ref{fig_PD}e. The PD patterns of needle-cable samples under different trigger voltage levels exhibited a similar PD distribution as needle-plate samples while requiring significantly higher applied voltage levels for each stage. This can be attributed to the reason that cable geometry trapped fewer surface charges with significantly smaller active charge distribution surface area, leads to lower surface potential.

\section{Conclusion}
A back discharge movement anomaly was found by applying periodic negative half-sine voltage waveform with a 0 V relaxation time period under DBCD experimental setups. The concentration of back discharge cluster during the relaxation time period, which derives from accumulated surface charges, moves with different voltage levels, manifesting some regularities with respect to PD amplitude, air gap length, insulation geometry, and thickness. Ionic wind theory is introduced to explain this phenomenon. Some conclusions are drawn as follows:

1) Back discharge under negative excitation presents significantly different PD patterns from one under positive excitation, as different mechanisms of back discharge were relied on.

2) Average PD amplitude shows an inverse relationship with voltage levels, as higher voltage generates a stronger ionic wind. Materials with higher resistivity also tend to yield lower back discharge amplitude.

3) The thickness of the insulation material has an influence on the surface charge density and surface potential, further affecting the back discharge amplitude as well as the voltage level for each stage.

4) Air gap length is a prerequisite of ionic wind propagation, different air gap lengths lead to different back discharge amplitude. In addition, ionic wind diminishes in narrow air gap length, with streamer discharge coming to the dominant discharge mechanism.

The analyses in this study are mainly based on the observation results of an intriguing back discharge movement phenomenon, which is influenced by the combination action of ionic wind and space charge distribution. Although It is always hard to analyze this type of multi-source phenomenon quantitatively, it is still helpful for understanding the main acting mechanism of DBCD and space charge dynamic. Nevertheless, statistical and visualized surface charge distribution and surface potential distribution measurements are expected in future work.


\bibliographystyle{ieeetr}
\bibliography{Ionic_wind/references}

@article{2yehia_characteristics_2019,
	title = {Characteristics of the dielectric barrier corona discharges},
	volume = {9},
	issn = {2158-3226},
	url = {https://pubs.aip.org/adv/article/9/4/045214/1076526/Characteristics-of-the-dielectric-barrier-corona},
	doi = {10.1063/1.5085675},
	pages = {045214},
	number = {4},
	journaltitle = {{AIP} Advances},
	author = {Yehia, Ashraf},
	urldate = {2023-08-15},
	date = {2019-04-01},
	langid = {english},
}

@inproceedings{13morsalin_corona_2018,
	location = {Xi'an},
	title = {Corona discharge under non-sinusoidal voltage excitation at very low frequency},
	isbn = {978-1-5386-5788-1},
	url = {https://ieeexplore.ieee.org/document/8401102/},
	doi = {10.1109/ICPADM.2018.8401102},
	eventtitle = {2018 12th International Conference on the Properties and Applications of Dielectric Materials ({ICPADM})},
	pages = {653--656},
	booktitle = {2018 12th International Conference on the Properties and Applications of Dielectric Materials ({ICPADM})},
	publisher = {{IEEE}},
	author = {Morsalin, S. and Phung, B. T.},
	urldate = {2023-08-15},
	date = {2018-05},
	langid = {english},
}

@article{6hammarstrom,
	title = {Detection of Electrical Tree Formation in {XLPE} Insulation through Applying Disturbed {DC} Waveforms},
	volume = {28},
	issn = {1070-9878, 1558-4135},
	url = {https://ieeexplore.ieee.org/document/9594886/},
	doi = {10.1109/TDEI.2021.009717},
	pages = {1669--1676},
	number = {5},
	journaltitle = {{IEEE} Transactions on Dielectrics and Electrical Insulation},
	shortjournal = {{IEEE} Trans. Dielect. Electr. Insul.},
	author = {Hammarstrom, Thomas and Gubanski, Stanislaw M.},
	urldate = {2023-08-15},
	date = {2021-10},
	langid = {english},
}

@book{25beroual_discharge_2016,
	location = {Bristol [England] (Temple Circus, Temple Way, Bristol {BS}1 6HG, {UK})},
	title = {Discharge in long air gaps: modelling and applications},
	isbn = {978-0-7503-1236-3},
	shorttitle = {Discharge in long air gaps},
	publisher = {{IOP} Publishing},
	author = {Beroual, A. and Fofana, I.},
	date = {2016},
	langid = {english},
	note = {{OCLC}: 953228826},
}

@article{8wu,
	title = {Effects of discharge area and surface conductivity on partial discharge behavior in voids under square voltages},
	volume = {14},
	issn = {1070-9878},
	url = {http://ieeexplore.ieee.org/document/4150615/},
	doi = {10.1109/TDEI.2007.344627},
	pages = {461--470},
	number = {2},
	journaltitle = {{IEEE} Transactions on Dielectrics and Electrical Insulation},
	shortjournal = {{IEEE} Trans. Dielect. Electr. Insul.},
	author = {Wu, Kai and Okamoto, Tatsuki and Suzuoki, Yasuo},
	urldate = {2023-08-15},
	date = {2007-04},
	langid = {english},
}

@article{14wang,
	title = {Enhanced distinction of surface and cavity discharges by trapezoid-based arbitrary voltage waveforms},
	volume = {23},
	issn = {1070-9878},
	url = {http://ieeexplore.ieee.org/document/7422589/},
	doi = {10.1109/TDEI.2015.005236},
	pages = {435--443},
	number = {1},
	journaltitle = {{IEEE} Transactions on Dielectrics and Electrical Insulation},
	shortjournal = {{IEEE} Trans. Dielect. Electr. Insul.},
	author = {Wang, X. and Taylor, N. and Edin, H.},
	urldate = {2023-08-15},
	date = {2016-02},
	langid = {english},
}

@article{5lindell,
	title = {Influence of rise time on partial discharge extinction voltage at semi-square voltage waveforms},
	volume = {17},
	issn = {1070-9878},
	url = {http://ieeexplore.ieee.org/document/5412012/},
	doi = {10.1109/TDEI.2010.5412012},
	pages = {141--148},
	number = {1},
	journaltitle = {{IEEE} Transactions on Dielectrics and Electrical Insulation},
	shortjournal = {{IEEE} Trans. Dielect. Electr. Insul.},
	author = {Lindell, Elisabeth and Bengtsson, Tord and Blennow, Jorgen and Gubanski, Stanislaw},
	urldate = {2023-08-15},
	date = {2010-02},
	langid = {english},
}

@article{3martins_modeling_2011,
	title = {Modeling of an {EHD} corona flow in nitrogen gas using an asymmetric capacitor for propulsion},
	volume = {69},
	issn = {03043886},
	url = {https://linkinghub.elsevier.com/retrieve/pii/S0304388611000258},
	doi = {10.1016/j.elstat.2011.02.002},
	pages = {133--138},
	number = {2},
	journaltitle = {Journal of Electrostatics},
	shortjournal = {Journal of Electrostatics},
	author = {Martins, Alexandre A. and Pinheiro, Mario J.},
	urldate = {2023-08-15},
	date = {2011-04},
	langid = {english},
}

@inproceedings{10wang,
	location = {West Lafayette, {IN}},
	title = {Partial discharge analysis of a narrow dielectric gap with repetitive half-sine pulses},
	isbn = {978-1-4244-9468-2},
	url = {http://ieeexplore.ieee.org/document/5724053/},
	doi = {10.1109/CEIDP.2010.5724053},
	eventtitle = {2010 {IEEE} Conference on Electrical Insulation and Dielectric Phenomena ({CEIDP} 2010)},
	pages = {1--4},
	booktitle = {2010 Annual Report Conference on Electrical Insulation and Dielectic Phenomena},
	publisher = {{IEEE}},
	author = {Wang, X L and Clemence, R and Edin, H},
	urldate = {2023-08-15},
	date = {2010-10},
	langid = {english},
}

@inproceedings{19yin_simulation_2016,
	location = {Xi'an, China},
	title = {Simulation and laboratory investigation of ionic wind induced by corona discharge},
	isbn = {978-1-5090-5418-3},
	url = {http://ieeexplore.ieee.org/document/7779550/},
	doi = {10.1109/APPEEC.2016.7779550},
	eventtitle = {2016 {IEEE} {PES} Asia-Pacific Power and Energy Engineering Conference ({APPEEC})},
	pages = {478--481},
	booktitle = {2016 {IEEE} {PES} Asia-Pacific Power and Energy Engineering Conference ({APPEEC})},
	publisher = {{IEEE}},
	author = {Yin, Fanghui and Farzaneh, Masoud and {Xingliang Jiang}},
	urldate = {2023-08-15},
	date = {2016-10},
	langid = {english},
}

@article{24fofana_study_2008,
	title = {Study of discharge in air from the tip of an icicle},
	volume = {15},
	issn = {1070-9878},
	url = {http://ieeexplore.ieee.org/document/4543110/},
	doi = {10.1109/TDEI.2008.4543110},
	pages = {730--740},
	number = {3},
	journaltitle = {{IEEE} Transactions on Dielectrics and Electrical Insulation},
	shortjournal = {{IEEE} Trans. Dielect. Electr. Insul.},
	author = {Fofana, I. and Farzaneh, M. and Hemmatjou, H. and Volat, C.},
	urldate = {2023-08-15},
	date = {2008-06},
	langid = {english},
}

@article{22goldman_corona_1985,
	title = {The corona discharge, its properties and specific uses},
	volume = {57},
	issn = {1365-3075, 0033-4545},
	url = {https://www.degruyter.com/document/doi/10.1351/pac198557091353/html},
	doi = {10.1351/pac198557091353},
	pages = {1353--1362},
	number = {9},
	journaltitle = {Pure and Applied Chemistry},
	author = {Goldman, M. and Goldman, A. and Sigmond, R. S.},
	urldate = {2023-08-15},
	date = {1985-01-01},
	langid = {english},
}

@article{1timatkov_influence_2005,
	title = {Influence of solid dielectric on the impulse discharge behaviour in a needle-to-plane air gap},
	volume = {38},
	issn = {0022-3727, 1361-6463},
	url = {https://iopscience.iop.org/article/10.1088/0022-3727/38/6/016},
	doi = {10.1088/0022-3727/38/6/016},
	pages = {877--886},
	number = {6},
	journaltitle = {Journal of Physics D: Applied Physics},
	shortjournal = {J. Phys. D: Appl. Phys.},
	author = {Timatkov, V V and Pietsch, G J and Saveliev, A B and Sokolova, M V and Temnikov, A G},
	urldate = {2023-08-15},
	date = {2005-03-21},
	langid = {english},
}

@article{4romano,
	title = {A New Approach to Partial Discharge Detection Under {DC} Voltage: Application to Different Materials},
	volume = {37},
	issn = {0883-7554, 1558-4402},
	url = {https://ieeexplore.ieee.org/document/9352713/},
	doi = {10.1109/MEI.2021.9352713},
	shorttitle = {A New Approach to Partial Discharge Detection Under {DC} Voltage},
	pages = {18--32},
	number = {2},
	journaltitle = {{IEEE} Electrical Insulation Magazine},
	shortjournal = {{IEEE} Electr. Insul. Mag.},
	author = {Romano, Pietro and Imburgia, Antonino and Rizzo, Giuseppe and Ala, Guido and Candela, Roberto},
	urldate = {2023-08-15},
	date = {2021-03},
	langid = {english},
}

@inproceedings{9guastavino,
	location = {Nashville, {TN}, {USA}},
	title = {A study about partial discharge measurements performed applying to insulating systems square voltages with different rise times},
	isbn = {978-0-7803-9257-1},
	url = {http://ieeexplore.ieee.org/document/1560709/},
	doi = {10.1109/CEIDP.2005.1560709},
	eventtitle = {{CEIDP} '05. 2005 Annual Report Conference on Electrical Insulation and Dielectric Phenomena, 2005.},
	pages = {418--421},
	booktitle = {{CEIDP} '05. 2005 Annual Report Conference on Electrical Insulation and Dielectric Phenomena, 2005.},
	publisher = {{IEEE}},
	author = {Guastavino, F. and Coletti, G. and Ratto, A. and Torello, E.},
	urldate = {2023-08-15},
	date = {2005},
	langid = {english},
}

@article{17soloviev_analytical_2012,
	title = {Analytical estimation of the thrust generated by a surface dielectric barrier discharge},
	volume = {45},
	issn = {0022-3727, 1361-6463},
	url = {https://iopscience.iop.org/article/10.1088/0022-3727/45/2/025205},
	doi = {10.1088/0022-3727/45/2/025205},
	pages = {025205},
	number = {2},
	journaltitle = {Journal of Physics D: Applied Physics},
	shortjournal = {J. Phys. D: Appl. Phys.},
	author = {Soloviev, V R},
	urldate = {2023-08-15},
	date = {2012-01-18},
	langid = {english},
}

@article{21shimizu_basic_2015,
	title = {Basic Study on Flow Control by Using Plasma Actuator},
	volume = {51},
	issn = {0093-9994, 1939-9367},
	url = {http://ieeexplore.ieee.org/document/7027789/},
	doi = {10.1109/TIA.2015.2397174},
	pages = {3472--3478},
	number = {4},
	journaltitle = {{IEEE} Transactions on Industry Applications},
	shortjournal = {{IEEE} Trans. on Ind. Applicat.},
	author = {Shimizu, Kazuo and Mizuno, Yoshinori and Blajan, Marius},
	urldate = {2023-08-15},
	date = {2015-07},
	langid = {english},
}

@article{7hammarstrom_combination_2021,
	title = {Combination of Adjustable Inverter Level and Voltage Rise Time for Electrical Stress Reduction in {PWM} Driven Motor Windings},
	volume = {37},
	issn = {0883-7554, 1558-4402},
	url = {https://ieeexplore.ieee.org/document/9290459/},
	doi = {10.1109/MEI.2021.9290459},
	pages = {17--26},
	number = {1},
	journaltitle = {{IEEE} Electrical Insulation Magazine},
	shortjournal = {{IEEE} Electr. Insul. Mag.},
	author = {Hammarstrom, T. J. A.},
	urldate = {2023-08-15},
	date = {2021-01},
	langid = {english},
}

@article{15gui_partial_2020,
	title = {Partial discharge characteristics of an air gap defect in the epoxy resin of a saturable reactor under an exponential decay pulse voltage},
	volume = {5},
	issn = {2397-7264, 2397-7264},
	url = {https://onlinelibrary.wiley.com/doi/10.1049/hve.2019.0121},
	doi = {10.1049/hve.2019.0121},
	pages = {482--488},
	number = {4},
	journaltitle = {High Voltage},
	shortjournal = {High Voltage},
	author = {Gui, Lu and Mi, Yan and Deng, Shengchu and Liu, Lulu and Ge, Xin and Ouyang, Wenmin},
	urldate = {2023-08-15},
	date = {2020-08},
	langid = {english},
}

@article{16article,
author = {Tajmar, Martin},
year = {2004},
month = {02},
pages = {315-318},
title = {Biefeld-Brown Effect: Misinterpretation of Corona Wind Phenomena},
volume = {42},
journal = {Aiaa Journal - AIAA J},
doi = {10.2514/1.9095}
}

@article{11suwarno_partial_1996,
	title = {Partial discharges due to electrical treeing in polymers: phase-resolved and time-sequence observation and analysis},
	volume = {29},
	issn = {0022-3727, 1361-6463},
	url = {https://iopscience.iop.org/article/10.1088/0022-3727/29/11/028},
	doi = {10.1088/0022-3727/29/11/028},
	shorttitle = {Partial discharges due to electrical treeing in polymers},
	pages = {2922--2931},
	number = {11},
	journaltitle = {Journal of Physics D: Applied Physics},
	shortjournal = {J. Phys. D: Appl. Phys.},
	author = {{Suwarno} and Suzuoki, Y and Komori, F and Mizutani, T},
	urldate = {2023-08-15},
	date = {1996-11-14},
	langid = {english},
}

@inproceedings{20mehmood_analysis_2019,
	location = {Taxila, Pakistan},
	title = {Analysis on the Propulsion of Ionic Wind During Corona Discharge in Various Electrode Configuration with High Voltage Sources},
	isbn = {978-1-72812-353-0},
	url = {https://ieeexplore.ieee.org/document/8853661/},
	doi = {10.1109/ICAEM.2019.8853661},
	eventtitle = {2019 International Conference on Applied and Engineering Mathematics ({ICAEM})},
	pages = {7--12},
	booktitle = {2019 International Conference on Applied and Engineering Mathematics ({ICAEM})},
	publisher = {{IEEE}},
	author = {Mehmood, Aqib and Jamal, Hassan},
	urldate = {2023-08-15},
	date = {2019-08},
	langid = {english},
}

@article{12mizutani2006pulse,
  title={Pulse-Sequence Analysis of Discharges in Air, Liquid and Solid Insulating Materials},
  author={Mizutani, Teruyoshi and others},
  journal={Journal of Electrical Engineering \& Technology},
  volume={1},
  number={4},
  pages={528--533},
  year={2006}
}

@inproceedings{18kitahara_experimental_2007,
	location = {Vancouver, {BC}, Canada},
	title = {An experimental analysis of ionic wind velocity characteristics in a needle-plate electrode system by means of laser-induced phosphorescence},
	isbn = {978-1-4244-1481-9},
	url = {http://ieeexplore.ieee.org/document/4451475/},
	doi = {10.1109/CEIDP.2007.4451475},
	eventtitle = {2007 Annual Report - Conference on Electrical Insulation and Dielectric Phenomena},
	pages = {529--532},
	booktitle = {2007 Annual Report - Conference on Electrical Insulation and Dielectric Phenomena},
	publisher = {{IEEE}},
	author = {Kitahara, Y. and Aoyagi, K. and Ohyama, R.},
	urldate = {2023-08-15},
	date = {2007},
	langid = {english},
}

@inproceedings{26wang2013lumped,
  title={Lumped-circuit Modeling of Surface Charge Decay in a Needle-plane geometry},
  author={Wang, Xiaolei and Taylor, Nathaniel and Niasar, M Ghaffarian and Kiiza, R Clemence and Edin, Hans},
  booktitle={Proceedings of the Nordic Insulation Symposium},
  number={23},
  year={2013}
}

@article{24morrowTheoryPositiveGlow1997a,
	title = {The theory of positive glow corona},
	volume = {30},
	issn = {0022-3727, 1361-6463},
	url = {https://iopscience.iop.org/article/10.1088/0022-3727/30/22/008},
	doi = {10.1088/0022-3727/30/22/008},
	pages = {3099--3114},
	number = {22},
	journaltitle = {Journal of Physics D: Applied Physics},
	shortjournal = {J. Phys. D: Appl. Phys.},
	author = {Morrow, R},
	urldate = {2023-08-29},
	date = {1997-11-21},
	langid = {english},
}

@ARTICLE{8785920,
  author={Meyer, Hans Kristian and Mauseth, Frank and Marskar, Robert and Pedersen, Atle and Blaszczyk, Andreas},
  journal={IEEE Transactions on Dielectrics and Electrical Insulation}, 
  title={Streamer and surface charge dynamics in non-uniform air gaps with a dielectric barrier}, 
  year={2019},
  volume={26},
  number={4},
  pages={1163-1171},
  doi={10.1109/TDEI.2019.007929}}

@article{guanMechanismsSurfaceCharge2019,
	title = {Mechanisms of surface charge dissipation of silicone rubber enhanced by dielectric barrier discharge plasma treatments},
	volume = {126},
	issn = {0021-8979, 1089-7550},
	url = {https://pubs.aip.org/jap/article/126/9/093301/280210/Mechanisms-of-surface-charge-dissipation-of},
	doi = {10.1063/1.5110615},
	language = {en},
	number = {9},
	urldate = {2023-08-31},
	journal = {Journal of Applied Physics},
	author = {Guan, Honglu and Chen, Xiangrong and Du, Hao and Paramane, Ashish and Zhou, Hao},
	month = sep,
	year = {2019},
	pages = {093301},
}

@article{fuPartialDischargeEvolution2022,
  title = {Partial {{Discharge Evolution}} under {{Half-sine Voltage Excitation}}},
  author = {Fu, Gan and Edin, Hans and Durga Pawan Gorla, Mahidhar and Janus, Patrick},
  date = {2022-07-05},
  journaltitle = {Proceedings of the Nordic Insulation Symposium},
  shortjournal = {NORD-IS},
  volume = {27},
  number = {1},
  issn = {2535-3969},
  doi = {10.5324/nordis.v27i1.4702}
}




\end{document}